\documentclass[]{raa}
\usepackage{amssymb,amsmath}            
\usepackage{graphicx,times}
\usepackage{natbib}
\usepackage{color}

\providecommand{\bm}[1]{\mbox{\bol
dmath$#1$\unboldmath}}

\begin{document}

   \title{Population I Cepheids and understanding star formation history of the Small Magellanic Cloud}

 \volnopage{ {\bf 2015} Vol.\ {\bf X} No. {\bf XX}, 000--000}
   \setcounter{page}{1}

   \author{Y. C. Joshi\inst{1}, A. P. Mohanty\inst{2}
\thanks{email: \texttt{aurogreenindia@gmail.com}},
 S. Joshi\inst{1} }

   \institute{ Aryabhatta Research Institute of Observational Sciences, Manora Peak, Nainital, India - 263002\\
        \and
             National Institute of Technology, Rourkela, India - 769008\\
\vs \no
   {\small Received 22 August 2015 ; Accepted 16 October 2015}
}

\abstract{In this paper, we study the age and spatial distributions of Cepheids in the Small Magellanic Cloud (SMC) as a function of their ages using the data from the OGLE III photometric catalogue. A period - age (PA) relation derived for the Classical Cepheids in the Large Magellanic Cloud (LMC) has been used to find the ages of Cepheids. The age distribution of the SMC Classical Cepheids is found to have a peak at $\log (Age) = 8.40 \pm 0.10$ which suggests that a major star formation event might have occurred in the SMC at about $250 \pm 50$ Myrs ago. It is believed that this star forming burst had been triggered by close interactions of the SMC with the LMC and/or the Milky Way (MW). A comparison of the observed spatial distributions of the Cepheids and open star clusters has also been carried out to study the star formation scenario in the SMC.
}

   \authorrunning{Joshi et~al. }            
   \titlerunning{Star formation history of the SMC using CCs}  
   \maketitle

%
\section{Introduction}\label{s:intro}
The SMC is the closest satellite galaxy of the MW after the LMC which is about 60 kpc away from us and can be seen in the Southern hemisphere. The SMC is proximal enough to provide us with an excellent opportunity to study its star formation history thus helping us to know the epochs and activities that led to its formation. 
Star formation can be triggered by several mechanisms like a turbulent interstellar medium, self-induced gravitational collapse of the molecular cloud, tidal shocking, or cloud-cloud interactions \citep{Mck07}. In recent times, the distribution of stellar populations in the Magellanic Clouds (MCs) has been studied with variety of objects, e.g. star clusters \citep{Pie00, HZ09, Gla10}, Cepheid variables \citep{Alc99, Nik04, Jos14}, RR Lyrae variables \citep{Sub09, Has12, Wag13}, red clump stars \citep{Koe09, Sub12}, HI observations \citep{Sta04}, among others. These studies imply that the episodic star formation events have taken place in the MCs, most likely due to repeated interaction between the MCs and/or with the MW.

Population I Cepheids, also known as Classical Cepheids (CCs), have been widely used to reconstruct the history of star formation in the MCs because they are intrinsically bright, easily observable and ubiquitous. They are ideal objects to understand the star formation activity in the past 30 - 600 Myrs of the galaxies as a typical life of the CCs lies in this epoch. The light curves of CCs pulsating in the fundamental mode are of asymmetric nature with a steep rise to their amplitudes but with a slower fall. However, their first-overtone counterparts are more symmetric with much smaller amplitudes and shorter periods. Cepheids, as discovered by Leavitt (1912), obey a period-luminosity relation. Their pulsation period, colour, mass, and intrinsic luminosity are related to each other. This led to their immense uses in tracing young stellar populations and star forming regions in the extra-galactic systems \citep{Elm96}. In the past CCs have been employed to study the spatial structure of the MCs \citep{CC86, Has12, Jos14, Sub15}. In order to further understand the star formation history in the SMC, a study of the distribution of CCs with larger sample of data has been carried out in the present paper. We here aim to improve the understanding of the Cepheids age distribution of the SMC and present spatial map of the star formation in this dwarf galaxy through the period, age and spatial distributions of the CCs.

The paper is organized as follows: in Section~\ref{s:dat}, we give the information about the data used in the present analysis. The period, age and spatial distributions of Cepheids are discussed in Sections~\ref{s:pdc}, \ref{s:adc} and \ref{s:sdc}, respectively. A comparison of the distribution of Cepheids with the star clusters is made on Section~\ref{s:csc}. Our results are summarized in Section~\ref{s:sum}.

\section{Data}\label{s:dat}
Highly precise, and calibrated V-I photometric data of about 6.2 million stars were obtained from 41 fields in the SMC during the third phase of the Optical Gravitational Lensing Experiment \citep{Sos10}. The fields were observed between 2001 and 2008 in the survey covering about 14 square degrees in the sky using the 1.3-m Warsaw telescope at the Las Campanas Observatory, Chile. \citet {Uda08} contains the details of the reduction procedure, photometric calibration, and astrometric calibration. All photometric data of their survey including the variable stars are available to the astronomical community from the OGLE web archive\footnote{http://ogle.astrouw.edu.pl/}. An extensive catalogue of 4630 CCs containing 2626 Fundamental mode (F), 1644 first-overtone (1O), 83 second-overtone (2O), 59 double-mode F/1O, 215 double-mode 1O/2O, and 3 triple-mode Cepheids are reported by \citet{Sos10}. The present work deals with a sample of 4270 CCs that include 2626 F and 1644 1O Cepheids taken from the above mentioned catalogue.
%
\section{The period distribution of Cepheids}\label{s:pdc}
\begin{figure}
\includegraphics[width=13.8cm, height=6.0cm]{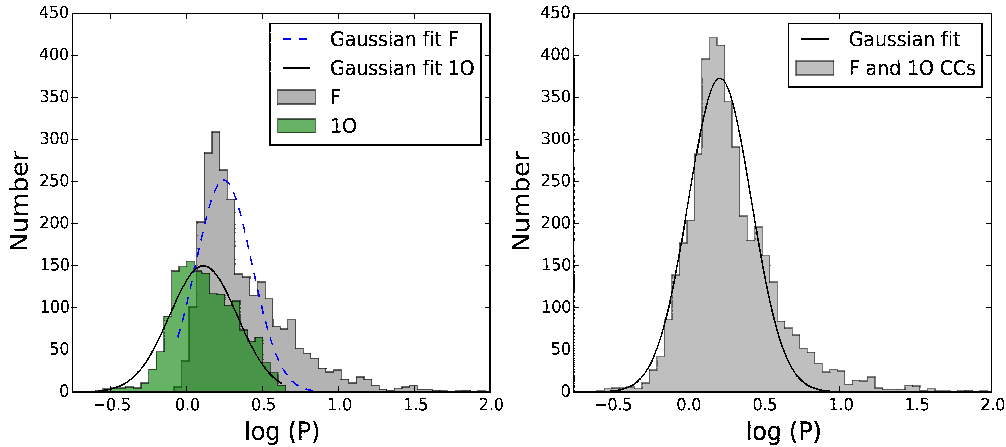}
\vspace{-0.5cm}
\caption{The period distribution of CCs in the SMC. Left panel shows both F and 1O Cepheids separately while right panel shows their combined distribution. The dotted blue and solid black lines show the best fit Gaussian profiles.}
\label{fig:pd}
\end{figure}

The periods, P (in days), of F CCs are in the range of $-0.08 < \log P < 2.31$, while that of 1O CCs lie in $-0.60 < \log P < 0.65$. The distribution of pulsation periods of CCs is different in different galaxies \citep {Jos10}. It also has peaks at different positions in the period distribution for the different galaxies. The period distribution depends upon chemical composition and many other factors like initial mass function, structure of the galaxy and the time spent by the stars pulsating during their transit through the instability strip \citep{Bec77, Alc99, Jos03}. We have drawn period distribution of the SMC CCs available in our sample with a bin width of $\Delta \log P = 0.05$ and shown in the right panel of Figure~\ref{fig:pd}. The individual period distributions of F and 1O CCs are shown in the left panel. As can be seen, there is a clear pattern in the distribution of CCs as a function of period. When we fit a Gaussian profile in the distribution, we obtain two peaks corresponding to these two class of CCs that lie at $ \log P = 0.25 \pm 0.01$ and $ \log P = 0.11 \pm 0.01$, for F and 1O CCs, respectively.

When we plot the period distribution of both the F and 1O CCs taken together, the combined sample gives an overall peak at $ \log P = 0.212 \pm 0.007$. It is quite evident that the Cepheids of longer periods are favoured over those with shorter periods. The period distribution for the SMC CCs deviates from the general pattern observed in the case of the LMC CCs as has been given by \citet {Jos14}.  Unlike SMC, they found two peaks in the LMC when both the F and 1O CCs were combined together, a result similar to the galaxies like M31 and the MW \citep{Antonello02, Macri04, Vilardell07, Jos10}. The absence of bimodal distribution in the case of the SMC could be due to the low metallicity of the SMC as chemical composition of the host galaxy may play a role, however small, in the period distribution of the Cepheids. A similar analysis done by \citet{Ser83}, \citet{Alc99}, and \citet{Jos10} in different galaxies also found that the frequency-period distribution varies in shape and in the location of the peak among different galaxies and is a function of the chemical composition.
%
\section{The age distribution of Cepheids}\label{s:adc}
There have been many successful attempts to derive empirical period - age (PA) relations for the Galactic, LMC, and M31 Cepheids. \citet{Efr78} gave an empirical period-age relation using the Cepheids in the Local Group of galaxies. \citet{Mag97} derived a semi-empirical period-age relation using Cepheids in the NGC~206 and superposed it in the M31 to trace the age distribution in order to understand the star formation history in that region. Later, \citet{Elf98} and \citet{Efr03} derived many PA relations considering different combinations of the 74 cluster Cepheids in 25 different open clusters in the LMC. 

\begin{figure}[ht]
\includegraphics[width=13.8 cm, height=6.0cm]{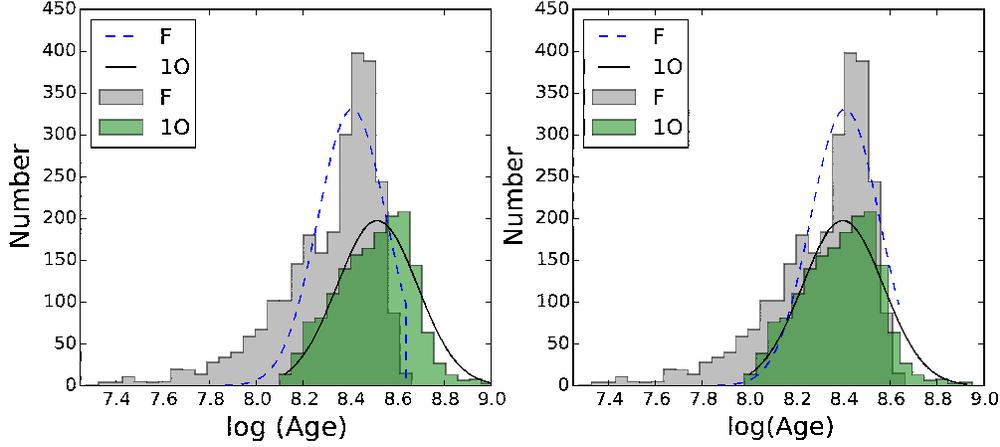}
\vspace{-0.5cm}
\caption{Age distribution of CCs in the SMC. The age distribution is shown in the left panel for F and 1O Cepheids. In the right panel, the age of 1O Cepheids are shown after the period conversion along with the F Cepheids. The dotted blue and solid black lines show the best fit Gaussian profiles.}
\label{fig:age}
\end{figure}

\citet{Jos14} made an attempt at improving the PA relation using the same sample of Cepheids given by the \citet{Efr03} but taking ages of the clusters from the more recent study of \citet{Pan10}. They drew a linear square fit to the points by plotting the mean period against the clusters age in logarithmic scale and derived the following relation:
\begin{equation}
\log (Age) = 8.60(\pm0.07) - 0.77(\pm0.08)~\log P
\end{equation}
Based on the evolutionary and pulsation models covering a broad range of stellar masses and chemical compositions, \citet{Bon05} derived a PA relation for the Cepheids with metallicity 0.004 as $\log (Age) = 8.49 - 0.79 \log P$. This is similar to the relation given by \citet{Jos14}. \citet{Bon05} did also present period-age-colour (PAC) relations apart from PA relations for F and 1O CCs. They found that the metal content of a galaxy affect the PA and PAC relations, though mildly. Short period Cepheids, which happen to be old, present minor intrinsic dispersions. Thus their ages could be more accurately estimated by using separate PA and PAC relations for Cepheids pulsating in the two modes. We however used the relation given by \citet{Jos14} to determine the ages of the CCs in the SMC. The F Cepheids in our sample have age ranges from 6.5 Myrs to 460.4 Myrs while the 1O Cepheids are between 125.2 Myrs and 1151.8 Myrs old. The age distribution of both F and 1O Cepheids are shown in the left panel of Figure~\ref{fig:age}. The age bin size in the distribution is $\Delta(\log(Age))=0.05$. We found two peaks occurring at $\log (Age) = 8.41$ and $\log (Age) = 8.52$ for F and 1O Cepheids, respectively.

\begin{figure}[h]
\begin{center}
\includegraphics[width=10.0 cm, height=5.5cm]{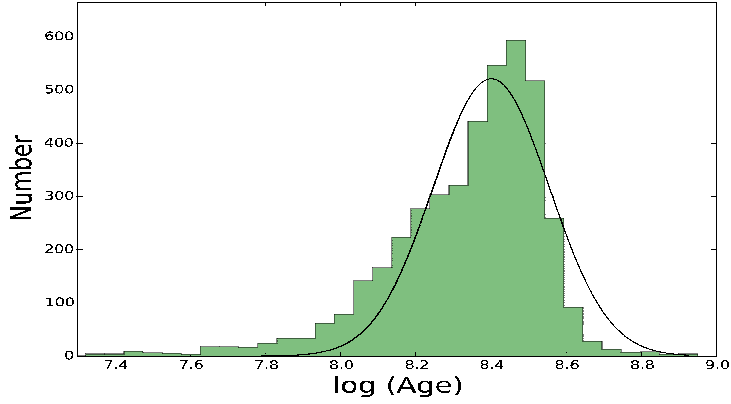}
\vspace{-0.5cm}
\caption{Combined age distribution of CCs after using the PA relation and converting the periods of 1O Cepheids to the corresponding F Cepheids. The solid black line shows the best fit Gaussian profile.}
\label{fig:comb}
\end{center}
\end{figure}

There exists a relation for transforming the periods of 1O Cepheids into the corresponding periods for the F Cepheids given by \citet{Alc95}. This empirical linear relation is ${P_1}/{P_0} = 0.733 - 0.034 \log P_1$, where $P_0$ and $P_1$ are periods of F and 1O Cepheids, respectively. However, \citet{Szi07} recently provided a more accurate linear relation from the spectroscopic observations of the Cepheids which is given as following.
\begin{equation}
P_1/P_0 = 0.710(\pm 0.001) - 0.014(\pm 0.003) \log P_0 - 0.027(\pm 0.004) \times [Fe/H]
\end{equation}
This relation is used to transform the periods of 1O Cepheids to that of F Cepheids. Here, we note that there is only a weak dependence of $[Fe/H]$ in the above conversion which can be ignored in the case of the SMC due to its extremely small metallicity. Following the above period conversion, the age distribution is drawn for F and 1O Cepheids separately in the right panel of Figure~\ref{fig:age}. The peak for 1O Cepheids is now shifted to log(Age) $\sim$ 8.40. If we combined both the sets of Cepheids after the period conversion, the overall peak of the age distribution occurs at $ \log (Age) = 8.40 \pm 0.01\pm0.09$ as shown in Figure~\ref{fig:comb}. Here, the first error corresponds to the statistical error in the mean age estimation in the Gaussian fit and second error represents the error due to uncertainty in the period-age relation. This suggests that there might be a major star formation event triggered in the SMC at around $250 \pm 50$ Myrs ago. When the two components of the MCs approach or recede, the star formation rate increases or decreases accordingly. Repeated tidal interaction between these two clouds thus leads to the episodic star formation events in both the dwarf galaxies. However, such star formation events in the MCs can also be induced through stellar winds and supernovae explosions through compression by turbulent motions \citep{Lar93, Gla10}.

\begin{figure}
\includegraphics[width=14.0 cm, height=9.5cm]{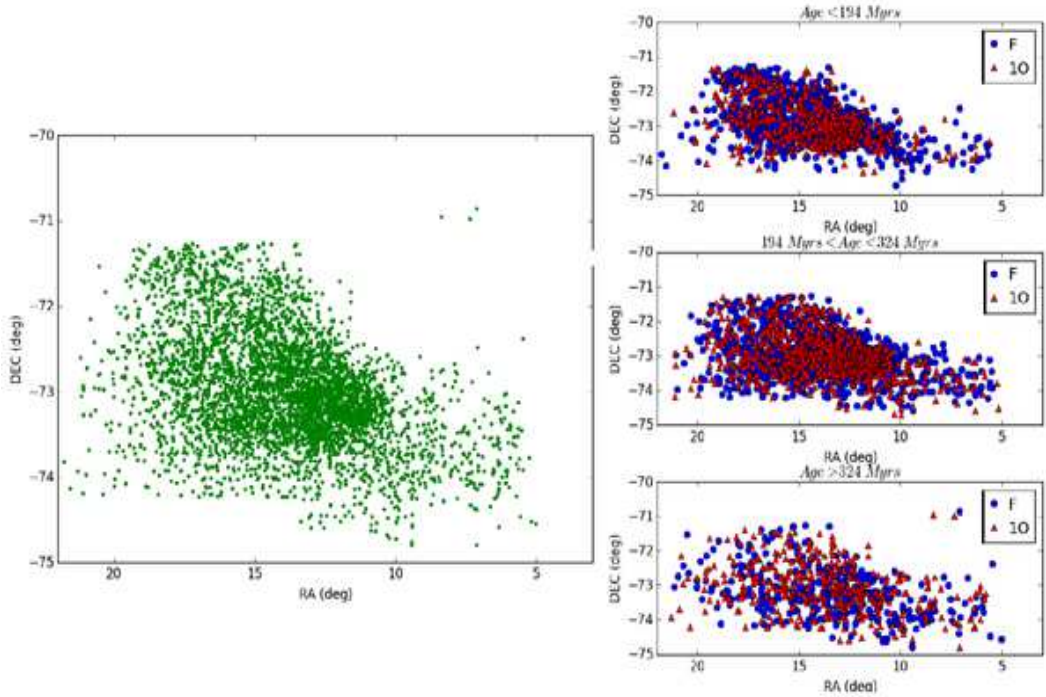}
\vspace{-0.5cm}
\caption{Spatial distribution of the CCs in the SMC. On the right panel, the distributions of F and 1O Cepheids for 3 different age intervals are shown with blue dots and red traingles, respectively.}
\label{fig:spatial}
\end{figure}

\section{The spatial distribution of Cepheids}\label{s:sdc}
To study the directional preference of the star formation in the SMC during the epoch of triggered star formation events, we analyse the spatial distribution of CCs and their age distributions within the SMC. From the distribution of Cepheids in the RA-DEC plane as shown in the left panel of Figure~\ref{fig:spatial}, it can be seen that there is a huge concentration of the Cepheids at the optical centre of the galaxy. In the right panels of the same figure, we draw similar distributions for Cepheids falling in the three different age intervals that are $200$ Myrs, $200 - 325$ Myrs, and $>325$ Myrs where F and 1O Cepheids are drawn with different colours. It is found that among all the F Cepheids, 36\% fall in the youngest age group, 51\% in the middle age group and remaining 13\% fall in the oldest age group. However, among all the 1O Cepheids, 30\% fall in the youngest category, 47\% in the middle age group and remaining 23\% in the oldest group. Although they are nearly in the same fraction over those different age groups it can be concluded that F Cepheids tend to be relatively younger.

\begin{figure}
\includegraphics[width=14.0 cm, height=9.5cm]{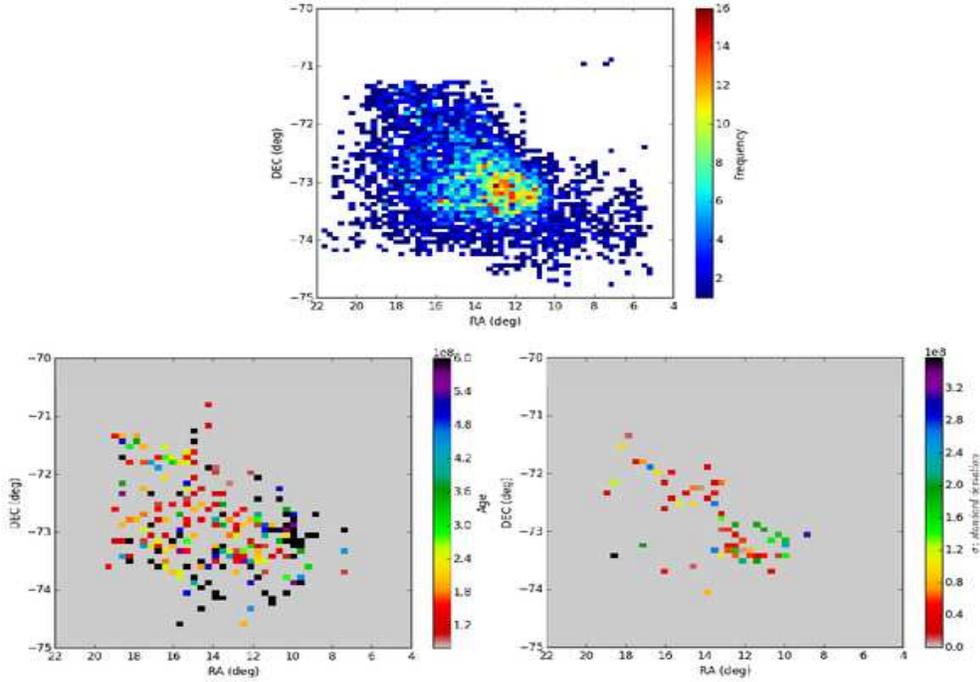}
\vspace{-0.5cm}
\caption{Map of CCs in the SMC in the RA-DEC plane. Top panel shows the spatial distribution. In the bottom panel, age distribution is shown on the left while a distribution of standard deviations is shown on the right.}
\label{fig:cpmap}
\end{figure}

\begin{figure}
\includegraphics[width=13.8cm]{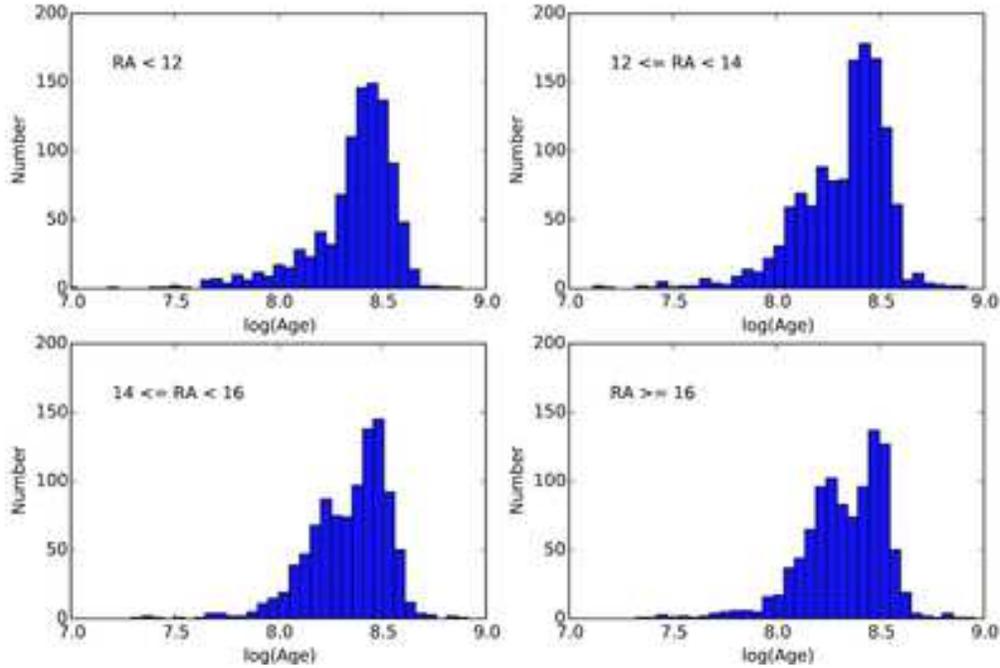}
\vspace{-0.3cm}
\caption{Age distribution of the Cepheids in four different regions of RA (deg) in the SMC as marked on the top of each panel.}
\label{fig:four}
\end{figure}

The star formation scenario has been studied in the radial direction of the SMC. The frequency distribution is shown on the top panel in Figure~\ref{fig:cpmap}. The RA-DEC plane is divided into 3600 (60 $\times$ 60) boxes. The distribution suggests that an explosion has triggered at the centre of the galaxy and stars are falling apart as the star forming event propagates. The age distribution shown at the bottom left panel and the dispersion in age of Cepheids shown at the bottom right panel further confirm this idea. There is a decrease in age along the diagonal towards the North-East which suggests that the star formation activity is propagating in this specified direction. There is less dispersion in ages on the border of the galaxy than towards the centre. This may suggest that star formation activities propagated from the centre to the outskirts of the galaxy.

Further, age distribution of Cepheids has been studied across the SMC by dividing them into four regions from West to East and shown in Figure~\ref{fig:four}. There is approximately equal number of Cepheids in these regions. The shape of the distribution in each of the regions is different but the peaks are approximately at the same age. It is normally found that the peak in the age distribution shifts towards larger value with the increase of metallicity. This suggests that there is zero metallicity gradient across the disk of the SMC. We can also clearly see that another peak emerges and gains significance as we proceed from West to East. This suggests that in those regions on the East, a separate star forming event might have occurred at about 160 Myrs ago (i.e. $\log(Age) = 8.2$) apart from the one major star forming burst at around 250 Myrs ago as discussed earlier. \citet{Pie00, Gla10} also found a peak at 160 Myr through the age distribution of star clusters in the SMC. The general profile based on the Cepheids age distribution matches well with that of star clusters in the SMC. From the analysis of CCs in the LMC, \citet{Jos14} has also noticed that there was a major star formation event triggered in the LMC at about 125-200 Myrs ago which was most likely triggered due to a close encounter between the SMC and the LMC. As this seems to be probably the same outburst which resulted simultaneously in both the dwarf galaxies due to an encounter between them, a combined analysis suggests that there was indeed a major star formation event that had happened in both the members of the MCs. According to models of \citet{BC05}, \citet{Kal06}, \citet{BKH12}, \citet{DB12}, and others, the last close encounters between the two components of the MCs had happened about 100-300 Myr ago and they show that the MW, the LMC and the SMC have interacted enough to produce the Magellanic stream between the clouds \citep{Gla10}.
\section{Spatial distribution of star clusters: a comparison}\label{s:csc}
It is quite interesting to make a comparison between the spatial distribution of star clusters and that of the Cepheids as the PA relation uses the age of clusters and the periods of Cepheids in finding the ages of the Cepheids. \citet{Pie99} plotted the age distribution of young clusters in the SMC and found two peaks at approximately 100 Myr and 160 Myr. \citet{Gla10} also found a pronounced peak at around 160 Myr for cluster formation in the SMC. From this, \citet{Sub15} shows that the age distribution of CCs and that of the clusters share a general profile.

\begin{figure}
\includegraphics[width=13.8 cm]{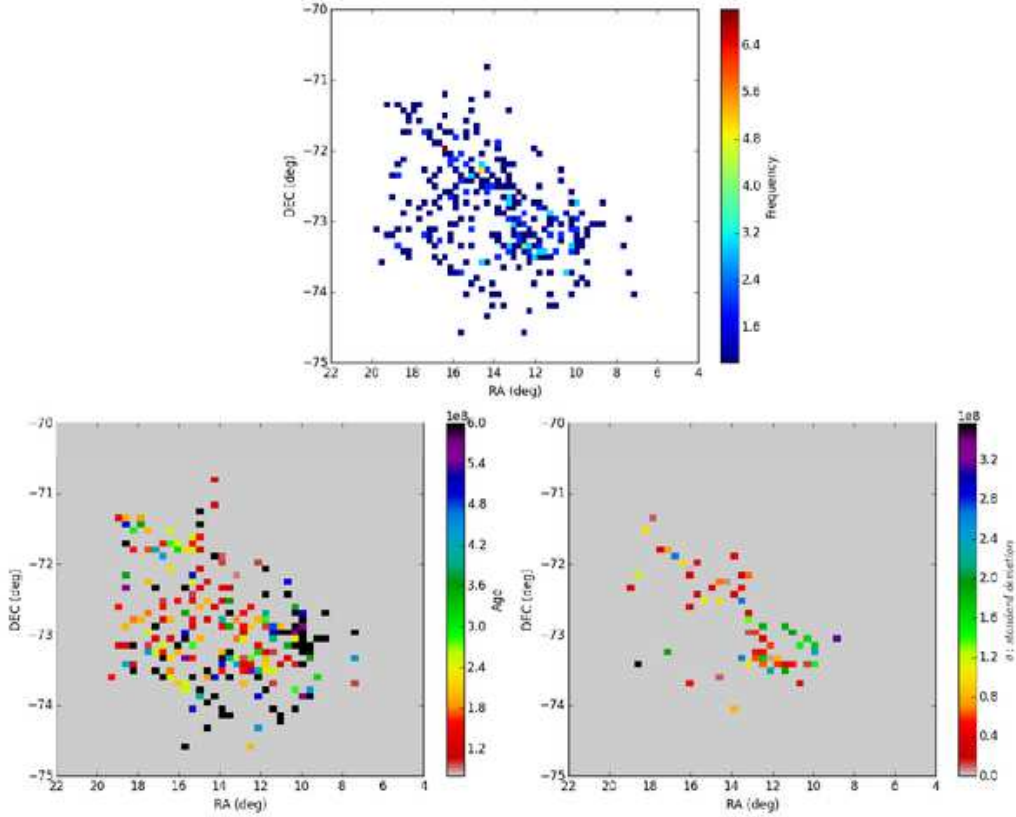}
\vspace{-0.3cm}
\caption {Map of open clusters in the SMC in the RA-DEC plane. Top panel shows the spatial distribution. In the bottom panel, age distribution is shown on the left and a distribution of standard deviations is shown on the right.}
\label{fig:clmap}
\end{figure}
\citet{Pie99} had given the ages of 93 star clusters in the SMC. \citet{Gla10} found the ages of 324 star clusters in the SMC, although four of them are duplicate clusters having different ages. This was found out while comparing the celestial coordinates of all the clusters given by \citet{Pie99} with those given by \citet{Gla10}. Therefore, we considered the mean of the two given age values for each of these four clusters. The frequency distribution for the clusters is shown in the top panel of Figure~\ref{fig:clmap}. The clusters are seen to be distributed along the SMC bar in an elongated structure. The region was divided into 2500 small boxes across the length of RA and DEC. With the same box size, average age of the clusters lying in each box was taken and an age distribution is plotted in the bottom left panel of Figure~\ref{fig:clmap}. The standard deviation of each box was also derived and the distribution is shown in the bottom right panel of Figure~\ref{fig:clmap}. It is seen from the frequency distribution of the clusters that their distribution is quite non-homogeneous and most of the clusters lie along the North-East South-West diagonal of the SMC and some spreading out of it. They seem to miss a circular region of half-a-degree radius on the East of the optical centre and they are scarcely populated around the boundaries. On comparing this distribution with the frequency distribution map of the Cepheids, it can be concluded that these star clusters tend to contain more of the younger stars. The age map of the clusters shows that older clusters lie in the South-West region while sample becomes younger as we move towards the North-East. There is minimal dispersion in ages of the clusters towards the North-East than the other regions. Thus dispersion is found to be related with the propagating star forming event.

\section{Summary}\label{s:sum}

Using the OGLE catalogue, we statistically analyse the star formation scenario in the SMC from very accurate period determinations of 2626 F and 1644 1O Cepheids detected in their third phase of observations. We have studied the period distribution of CCs and found a peak at $\log P = 0.212 \pm 0.007$. On plotting the period distributions of F and 1O Cepheids separately we yield peaks at $\log P = 0.25 \pm0.01$ and $\log P = 0.11 \pm 0.01$, respectively. On combining these two classes of pulsating stars, after converting the periods of 1O Cepheids to that of F Cepheids and employing a period-age relation for the LMC, we found an age distribution comprising a pronounced peak at $\log(Age) =8.40 \pm 0.10$. There might have triggered a major star formation event at around $250 \pm 50$ Myrs ago in the SMC followed by one other such event in the Eastern region of the SMC at about 160 Myrs ago. A detailed study of different populations in the SMC \citep{Pie00, Sta04, HZ09, Gla10, Sub12, Has12, Wag13, Jos14, Sub15} as well as the simulations by various groups \citep{BC05, BKH12, DB12} suggested that the last close encounters in the MCs had happened at around 100-300 Myrs ago which has altered the star formation scenario in both the LMC and SMC. Hence our results for the SMC in the present paper are in broad agreement with these previous studies.

Our study shows that the Cepheids have non-homogeneous distribution like a bursting balloon and are highly concentrated at the optical centre of this dwarf galaxy. It seems that the close encounters between the two components of the MCs and/or in between the MCs and the MW has induced episodic star formation in these galaxies. It suggests that the MCs are not in a bound system. On a comparison of spatial distribution of CCs and star clusters, a mutual correlation was noticed in the LMC, however, number of known clusters in the SMC is still highly incomplete to make any firm conclusion.
\normalem
\begin{acknowledgements}
This publication makes use of data products from the OGLE archive. APM is thankful to the Indian Academy of Sciences (IASc), Bangalore for the financial assistance provided through the IAS-SRFP 2014. We also thank Brajesh Kumar and Ramkesh Yadav for their valuable inputs which has helped to improve this paper. YCJ acknowledges the grant received under the Indo-Russian project INT/RUS/RFBR/P-219 funded by Department of Science and Technology, New Delhi.

\end{acknowledgements}


\bibliographystyle{raa}
\bibliography{bibtex}

\begin{thebibliography}{37}
\providecommand\natexlab[1]{#1}
\providecommand\JournalTitle[1]{#1}

\bibitem[Alcock et al.(1995)]{Alc95} 
Alcock, C., Allsman, R.A., Axelrod, T.S., et al., 1995. AJ 109, 1653

\bibitem[Alcock et al.(1999)]{Alc99}
Alcock, C., Allsman, R.A., Alves, D.R., et al., 1999. AJ 117, 920

\bibitem[Antonello et al.(2002)]{Antonello02}
Antonello, E., Fugazza, D., \& Mantegazza, L., 2002, A\&A, 388, 477

\bibitem[Becker et al.(1977)]{Bec77}
Becker, S. A., Iben Jr., I., Tuggle, R. S., 1977. ApJ 218, 633-653

\bibitem[Bekki \& Chiba(2005)]{BC05}
Bekki, K., \& Chiba, M., 2005, MNRAS, 356, 680

\bibitem[Besla et al.(2012)]{BKH12}
Besla, G., Kallivayalil, N, Hernquist, L., et al., 2012, MNRAS, 421, 2109 

\bibitem[Bono et al.(2005)]{Bon05}
Bono, C., Marconi, M., Cassisi, S., et al., 2005. ApJ 621, 966

\bibitem[Caldwell \& Coulson(1986)]{CC86}
Caldwell, J. A.. R., \& Coulson, I. M., 1986, MNRAS, 218, 223

\bibitem[Diaz \& Bekki(2012)]{DB12}
Diaz, J. D., \& Bekki, K., 2012, ApJ, 750, 36

\bibitem[Elmegreen \& Efremov(1996)]{Elm96} 
Elmegreen, B.G., Efremov, Y.N., 1996. ApJ 466, 802

\bibitem[Efremov \& Elmegreen(1998)]{Elf98} 
Efremov, Y.N., Elmegreen, B.G., 1998. MNRAS 299, 588

\bibitem[Efremov(1978)]{Efr78}
Efremov, Y.N., 1978. Soviet Astron. 22, 161

\bibitem[Efremov(2003)]{Efr03} 
Efremov, Y.N., 2003. Astron. Rep. 47, 1000

\bibitem[Glatt et al.(2010)]{Gla10}
Glatt, K., Grebel, E.K., Koch, A., 2010. A\&A 517, 50

\bibitem[Harris \& Zaritsky(2009)]{HZ09}
Harris, J., \& Zaritsky, D., 2009, AJ, 138, 1243

\bibitem[Haschke et al.(2012)]{Has12}
Haschke, R., Grebel, E. K., Duffau, S., 2012, AJ, 144, 107

\bibitem[Joshi et al.(2003)]{Jos03} 
Joshi, Y. C., Pandey, A. K., Narasimha, D., Sagar, R., Giraud-Héraud, Y., 2003, A\&A, 402, 113

\bibitem[Joshi et al.(2010)]{Jos10}
Joshi, Y.C., Narasimha, D., Pandey, A.K., Sagar, R., 2010. A\&A 512, 66

\bibitem[Joshi et al.(2014)]{Jos14}
Joshi, Y.C., Joshi, S., 2014. New Astronomy, 28, 27

\bibitem[Kallivayalil et al.(2006)]{Kal06}
Kallivayalil, N., van der Marel, R. P., \& Alcock, C., 2006, ApJ, 652, 1213

\bibitem[Koerwer(2009)]{Koe09}
Koerwer, J. F., 2009, AJ, 138, 1

\bibitem[Larson(1993)]{Lar93}
Larson, R. B., 1993, in The Globular Cluster-Galaxy Connection, ed. G. H. Smith \& J. P. Brodie, ASP Conf. Ser., 48, 675

\bibitem[Macri(2004)]{Macri04}
Macri, L. M., 2004, Variable Stars in the Local Group, ASP Conference Series, Vol. 2nn, 2004,
eds. D. W. Kurtz \& K. Pollard

\bibitem[Magnier et al.(1997)]{Mag97} 
Magnier, E.A., Augusteijn, T., Prins, S., van Paradijs, J., Lewin, W.H.G., 1997. A\&AS
126, 401

\bibitem[Mckee \& Ostriker(2007)]{Mck07} 
Mckee, C. F. \& Ostriker, E. C., 2007, ARA\&A, 45, 565

\bibitem[Nikoleav et al.(2004)]{Nik04}
Nikoleav, S., Drake, A. J., Keller, S. C., et al., 2004, ApJ, 601, 260

\bibitem[Pandey et al.(2010)]{Pan10} 
Pandey, A.K., Sandhu, T.S., Sagar, R., Battinelli, P., 2010. MNRAS 403, 1491

\bibitem[Pietrzynski \& Udalski(1999)]{Pie99}
Pietrzynski, G., \& Udalski, A., 1999. AcA 49, 157

\bibitem[Pietrzynski \& Udalski(2000)]{Pie00}
Pietrzynski, G. \& Udalski, A., 2000, A\&A, 50, 337

\bibitem[Serrano(1983)]{Ser83}
Serrano, A., 1983, RMxAA, 8, 131

\bibitem[Sziladi et al.(2007)]{Szi07}
Szil\'{a}di, K., Vink\'{o}, J., Poretti, E., Szabados, L., Kun, M., 2007. A\&A 473, 579

\bibitem[Soszynski et al.(2010)]{Sos10}
Soszynski, I., Poleski, R., Udalski, A., et al., 2010. AcA 60, 17

\bibitem[Stanimirovic et al.(2004)]{Sta04}
Stanimirovi\`{c}, S., \&  Staveley-Smith, L.; Jones, P. A., 2004, ApJ, 604, 176

\bibitem[Subramaniam \& Subramanian(2009)]{Sub09}
Subramaniam, A., \& Subramanian, I.,, 2009, A\&A, 503, 9

\bibitem[Subramanian \& Subramaniam(2012)]{Sub12}
Subramanian, I., \& Subramaniam, A., 2012, ApJ, 744, 128

\bibitem[Subramanian \& Subramaniam(2015)]{Sub15}
Subramanian, S., \& Subramaniam, A., 2015, A\&A, 573, 135

\bibitem[Udalski et al.(2008)]{Uda08}
Udalski, A., Szymanski, M.K., Soszynski, I., Poleski, R., 2008. AcA 58, 69

\bibitem[Vilardell et al.(2007)]{Vilardell07}
Vilardell, F., Jordi, C., \& Ribas, I., 2007, A\&A, 473, 847

\bibitem[Wagner-Kaiser(2013)]{Wag13}
Wagner-Kaiser, R., \& Sarajedini, A., 2013, MNRAS, 431, 1565

\end{thebibliography}


\label{lastpage}

\end{document}